\RequirePackage{lineno} 
\documentclass{nature}
\usepackage{xcolor}

\usepackage{caption}

\title{Spatially-resolved electronic structure of stripe domains in IrTe$_2$ through electronic structure microscopy}

\author{Changhua Bao$^1$, Hongyun Zhang$^1$, Qian Li$^1$, Shaohua Zhou$^1$, Haoxiong Zhang$^1$, Ke Deng$^1$, Kenan Zhang$^1$, Laipeng Luo$^1$,  Wei Yao$^1$, Chaoyu Chen$^2$, Jos\'{e} Avila$^2$, Maria C. Asensio$^{3}$, Yang Wu$^{4}$ \& Shuyun Zhou$^{1,5,*}$}

\usepackage{graphicx}
\makeatletter
\let\saved@includegraphics\includegraphics
\AtBeginDocument{\let\includegraphics\saved@includegraphics}

\makeatother

\begin{document}
\maketitle

\begin{affiliations}
 \item State Key Laboratory of Low-Dimensional Quantum Physics and Department of Physics, Tsinghua University, Beijing 100084, P. R. China
 \item Synchrotron SOLEIL, Universit\'e Paris-Saclay, L'Orme des Merisiers, Saint Aubin-BP 48, 91192 Gif-sur-Yvette Cedex, France
 \item Madrid Institute of Materials Science (ICMM), Spanish Scientific Research Council (CSIC), Cantoblanco, E-28049 Madrid, Spain
\item Department of Mechanical Engineering and Tsinghua-Foxconn Nanotechnology Research Center, Tsinghua University, Beijing 100084, P. R. China
\item Frontier Science Center for Quantum Information, Beijing 100084, P. R. China\\
 $^*$ Email: syzhou@mail.tsinghua.edu.cn
\end{affiliations}

\begin{abstract}
Abstract:\\
Phase separation in the nanometer- to micrometer-scale is characteristic for correlated materials, for example, high temperature superconductors, colossal magnetoresistance manganites, Mott insulators, etc. Resolving the electronic structure with spatially-resolved information is critical for revealing the fundamental physics of such inhomogeneous systems yet this is challenging  experimentally. Here by using nanometer- and micrometer-spot angle-resolved photoemission spectroscopies (NanoARPES and MicroARPES), we reveal the spatially-resolved electronic structure in the stripe phase of IrTe$_2$.  Each separated domain shows two-fold symmetric electronic structure  with the mirror axis aligned along 3 equivalent directions, and 6$\times$1 replicas are clearly identified. Moreover, such electronic structure inhomogeneity disappears across the stripe phase transition, suggesting that electronic phase with broken symmetry induced by the 6$\times$1 modulation  is directly related to the stripe phase transition of IrTe$_2$. Our work demonstrates the capability of NanoARPES and  MicroARPES in elucidating the fundamental physics of phase-separated materials.
\end{abstract}

\section*{Introduction}

By focusing the beam size down to a few $\mu$m or even 100 nm scale by a Fresnel zone plate\cite{Elettra, EliNanoARPES,AsensioNanoARPES} (for synchrotron light source) or a lens\cite{ShimadaMARPES} (for laser source), nanometer- and micrometer-spot angle-resolved photoemission spectroscopies (NanoARPES\cite{Elettra, EliNanoARPES, AsensioNanoARPES} and MicroARPES\cite{ShimadaMARPES}, Fig.~1a) provide two important advantages over conventional ARPES which has a typical beam size of 50-100 $\mu$m. Firstly, it allows to measure the electronic structure of small samples, which has been demonstrated in atomically thin flakes\cite{CobdenWSe2, MoTe2NanoARPES, JozwiakNP} or samples with mixed crystal orientations\cite{TrilayerG, tBLGPNAS}. Secondly and more importantly, for phase-separated materials which consist of multiple domains with distinct electronic structures\cite{CheongNat1999, DavisNat2001, DagottoSci2005, BasovSci2007, ShenSci2001}, the newly added spatial-resolving capability provides new opportunities to reveal the intrinsic electronic structure of individual domain and the evolution of the phase separation across the phase transition. Such information cannot be obtained by conventional ARPES, which is, however,  indispensable for understanding the fundamental physics of  phase-separated materials.
Recently, NanoARPES and MicroARPES have been applied to probe the electronic structure of individual domain in CeSb\cite{Kondo2020} and Fe-based superconductors\cite{Kim2019,Lanzara2020,Kim2020,KFeSe} by utilizing the spectroscopic capability of ARPES. Combining the advantages of both microscopic and spectroscopic capabilities of NanoARPES and MicroARPES will allow for direct visualization of separated domains with spatially-resolved information and the evolution of domains across the phase transition, thereby further elucidating the complex physics of phase-separated materials.

IrTe$_2$ exhibits an intriguing stripe phase with separated domains at low temperature, where the electronic structure in the stripe phase has remained elusive. Upon cooling, it undergoes a first order phase transition from trigonal (1T with P$\bar3$m1 symmetry, see Fig.~1b) to triclinic structure (P$\bar1$) around 280 K, accompanied by complex stripe phase\cite{Cheong2012} with periods of (3\textit{n}+2)$\times$1$\times$(3\textit{n}+2) (\textit{n} = 1,2,3,...) in the bulk\cite{Cheong2012, Kiryukhin2014} and (3\textit{n}+2)$\times$1 on the surface\cite{WuSTMPRL2013, Plummer2017}. Recent scanning tunnelling microscopy and ARPES study on strained IrTe$_2$ shows 6$\times$1 phase can be stabilized by strain\cite{Monney2021}. Suppressing the stripe phase by doping or intercalation\cite{Cheong2012, Ishiwata2013} leads to emergence of superconductivity. To understand the stripe phase, different scenarios have been proposed including Fermi surface nesting\cite{Cheong2012} or saddle point\cite{DH2014} induced charge density wave, crystal field effect\cite{Takashi2012}, dimerization\cite{PascutPRL2014, KimCheong2014}, local bonding\cite{David2013, Artyukhin2020} and lattice deformation\cite{Min2015}. Obtaining the electronic structure of the stripe phase is critical for disentangling the puzzling physics. Despite extensive investigations, previous ARPES measurements on IrTe$_2$\cite{ARPES2013,DH2014,Min2015,Park2017,ARPES2017,OOTSUKI2018,HengsbergerPRB2018,MonneyPRB2020} have been obtained by averaging over different domains, and the intrinsic electronic structure of each individual stripe domain and its temperature evolution across the phase transition remain elusive.

Here, by using NanoARPES and MicroARPES, we resolve the separated domains and electronic structure of individual stripe domain in IrTe$_2$. Each separated domain shows two-fold symmetric electronic structure with the mirror axis aligned along 3 equivalent directions, and 6$\times$1 reconstructions are clearly identified both in the Fermi surface map and the dispersion, suggesting 6$\times$1 stripe phase. Moreover, such electronic structure inhomogeneity disappears across the stripe phase transition, suggesting that electronic phase with broken symmetry induced by the 6$\times$1 modulation  is directly related to the stripe phase transition of IrTe$_2$. Our work demonstrates the power of NanoARPES and MicroARPES in elucidating the physics across the phase transition.

\section*{Results and discussion}

\textbf{Spatially-resolved electronic structures of different domains.} Figure 1d,e shows two representative NanoARPES spectra measured along the ${\bar{\Gamma}}$-${\bar K}$ direction (Fig.~1c) of IrTe$_2$ at 80 K from two domains A and B, and they are strikingly different. While the dispersion in domain A is relatively simple with strong intensity at energies from -2.5 to -0.5 eV, the dispersion in domain B shows weak intensity starting from -1 eV to the Fermi energy (E$\rm_F$) with many weaker bands near E$\rm_F$. In addition, compared to domain A, there is an additional band near the $\bar{K}$ point (marked by box 2 in Fig.~1e).  Spatially-resolved intensity maps (Fig.~1f,g) integrated over box 1 and 2 allow to directly visualize the spatial distribution of these two types of domains with size of a few to tens of micrometers. Since the sample is a high quality single crystal with a well-defined crystal orientation, the observation of  separated domains with different electronic structures therefore suggests that IrTe$_2$ is an intrinsically inhomogeneous material.

\begin{figure*}[htbp]
	\centering
	\includegraphics[width=16.8cm]{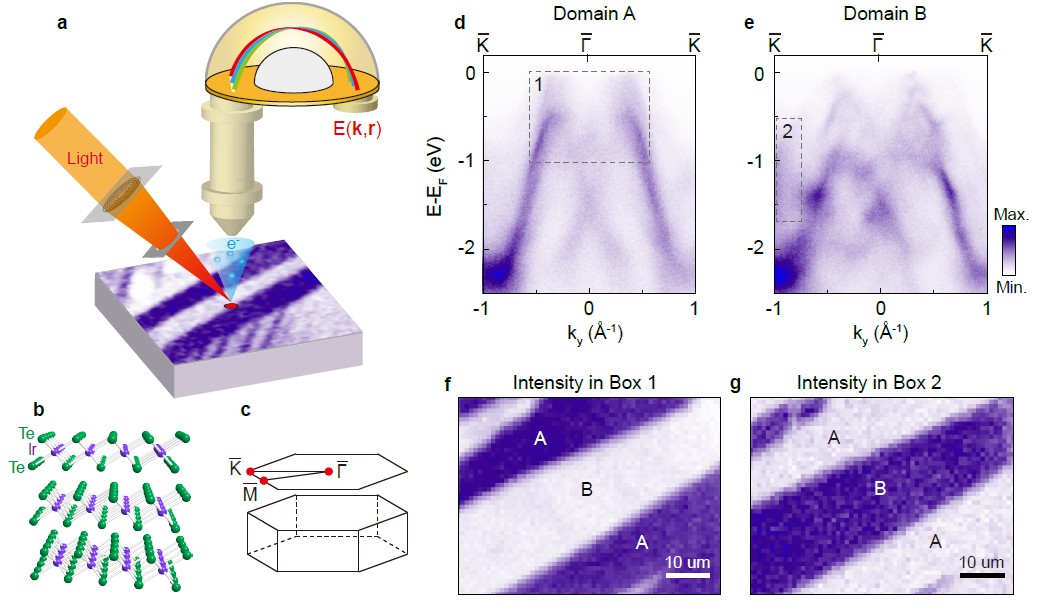}
	\caption{\textbf{Resolving the distinct electronic structures of separated domains in IrTe$_2$ by NanoARPES with energy-, momentum- and space-resolved information.} \textbf{a} A schematic drawing of NanoARPES and MicroARPES. The light beam is focused onto the sample by a Fresnel zone plate or a lens down to a few $\mu$m or even 100 nm scale and generate the photoelectrons which is measured by the analyzer. \textbf{b,c} Crystal structure of IrTe$_2$ and the corresponding Brillouin zone above the stripe transition temperature. \textbf{d,e} Two characteristic dispersions observed in domains A and B (as labeled in panels \textbf{f} and \textbf{g}) measured at photon energy of 100 eV with $p$-polarization and temperature of 80 K. \textbf{f} Spatially-resolved intensity map integrated over box 1 of panel \textbf{d}. \textbf{g} Spatially-resolved intensity map integrated over box 2 of panel \textbf{e}.}
	\label{Fig1}
\end{figure*}

\begin{figure*}[htbp]
	\centering
	\includegraphics[width=16.8 cm]{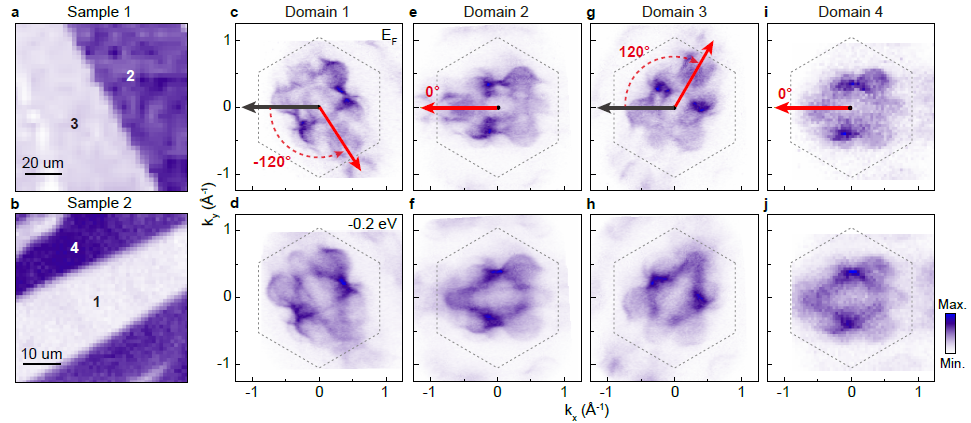}
	\caption{\textbf{Visualizing the stripe phases with three different orientations.} \textbf{a,b} Spatially-resolved intensity maps integrated over box 1 in Fig.~1d. \textbf{c-j} Fermi surface and intensity maps at -0.2 eV of corresponding domains in panels \textbf{a,b} with Brillouin zone of 1$\times$1 (gray hexagon).}
	\label{Fig2}
\end{figure*}

To further investigate the electronic structure of these separated domains, we map out the full three-dimensional electronic dispersions for each domain. Figure 2a,b shows the spatially-resolved intensity maps measured on two representative samples where separated domains are clearly observed. Figure 2c-j shows the intensity maps at $\rm E_F$ and -0.2 eV from four different domains. All these intensity maps clearly reveal the two-fold symmetry of the electronic structure with the symmetry axis aligned along three equivalent $\bar{\Gamma}$-$\bar{M}$ directions at angles of 0$^\circ$, 120$^\circ$ and -120$^\circ$ (indicated by red solid arrows), which is in sharp contrast to previous ARPES measurements\cite{ARPES2013,DH2014,Min2015,Park2017,ARPES2017,OOTSUKI2018,HengsbergerPRB2018,MonneyPRB2020} where spatial averaging gives rise to apparently three-fold symmetric electronic structure. Therefore, the strikingly different dispersions in Fig.~1 originate from different orientations of the mirror symmetry axes. Here, the observation of two-fold symmetric electronic structure in a three-fold symmetric crystal confirms the broken symmetry in the stripe phase, and the space- and momentum-resolving capability allows to reveal the intrinsic electronic structure of each individual domain.

\begin{figure*}[htbp]
	\centering
	\includegraphics[width =16.5 cm]{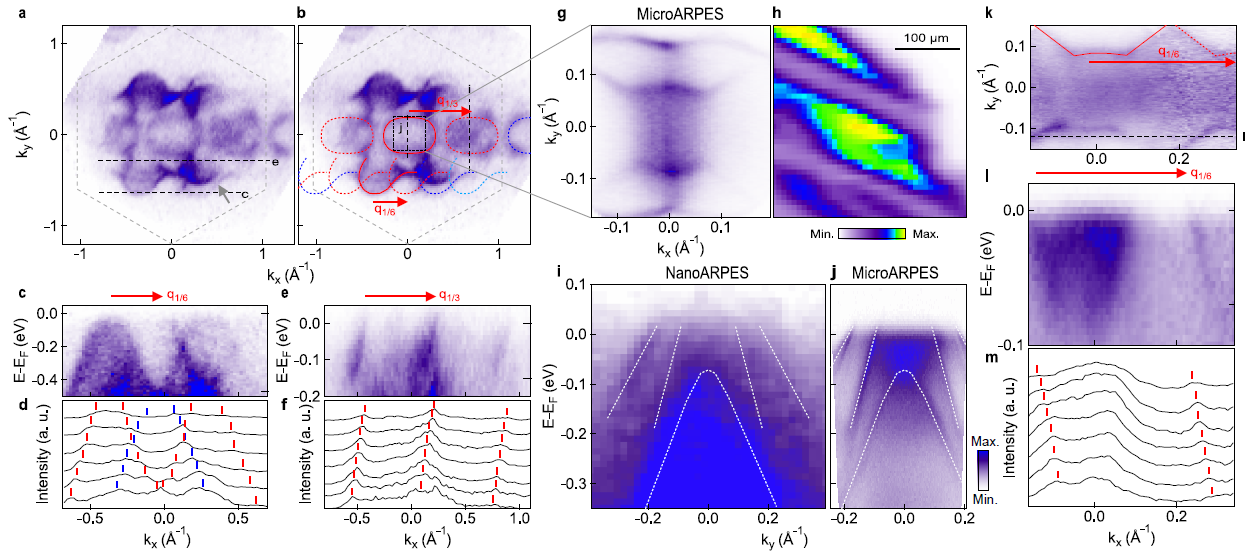}
	\caption{\textbf{Electronic reconstructions of 1/6 and 1/3 revealed by NanoARPES and MicroARPES.} \textbf{a,b} Fine Fermi surface maps measured by NanoARPES. Guides for the pockets (solid curves) and replicas (dotted curves) translated by \textbf{q}$_{1/3}$ and \textbf{q}$_{1/6}$ are overplotted in \textbf{b}. \textbf{c-f} Dispersion images (\textbf{c,e}) along the direction as indicated in  \textbf{a} and corresponding momentum distribution curves (MDC) (\textbf{d,f}).  \textbf{g} High-resolution zoom-in Fermi surface map by MicroARPES. \textbf{h} Spatially-resolved intensity map by integrating over momentum range of $k_y$=(-0.1 $\rm \AA^{-1}$,0.1 $\rm \AA^{-1}$) and energy range of $E$=(-0.05 $\rm eV$,0). \textbf{i,j} NanoARPES (\textbf{i}) and MicroARPES (\textbf{j}) dispersion images along the directions as indicated in \textbf{b}. The white curves are extracted dispersions from MicroARPES spectrum (\textbf{j}) and overplotted to NanoARPES spectrum in \textbf{i}. \textbf{k} Zoom-in Fermi surface map by MicroARPES. Solid curves are guides for the pockets and dotted curves are the replicas translated by \textbf{q}$_{1/6}$. \textbf{l,m} Dispersion images (\textbf{l}) along the direction as indicated in (\textbf{k}) and corresponding MDCs (\textbf{m}).}
	\label{Fig5}
\end{figure*}

With the capability to resolve the electronic structure of each individual domain, we can now investigate the intrinsic electronic structure and the nature of the stripe phase.
The two-fold symmetric Fermi surface map (Fig.~3a) shows replica oval pockets around the $\bar{\Gamma}$ point  translated by a scattering wave vectors of 1/3 \textbf{a}$^*$ (Fig.~3b) where \textbf{a}$^*$ is the reciprocal lattice vector. More replica pockets can be distinguished away from the $\bar{\Gamma}$ point translated by 1/6 \textbf{a}$^*$ which fit well with the extra weak bands as indicated by the gray arrow in Fig.~3a. The replica pockets are also identified in the dispersion images shown in Fig.~3c, e and can be observed more clearly in the momentum distribution curves (MDCs) shown in Fig.~3d, f.
We note that fine features are observed inside the replica oval pocket translated by 1/3 \textbf{a}$^*$ from $\bar{\Gamma}$. These features are absent in the oval pocket at  $\bar{\Gamma}$, yet their existence can be confirmed by zooming in the intensity map (Fig.~3g) near $\bar{\Gamma}$ point using
our home-built MicroARPES system with a laser source at 6.2 eV (compared to 100 eV used in synchrotron based NanoARPES measurements) with better energy and momentum resolution.  A comparison of NanoARPES and MicroARPES dispersion images (Fig.~3i,j) measured along two equivalent momentum directions displaced by 1/3 \textbf{a}$^*$ as indicated by dotted lines in Fig.~3b shows a good agreement yet with sharper peaks for MicroARPES, further confirming the electronic reconstruction with a scattering vector of 1/3 \textbf{a}$^*$.
The spatially-resolved intensity map measured by MicroARPES in Fig.~3h also shows separated domains with domain size up to hundred micrometers. The electronic reconstruction of 1/6 \textbf{a}$^*$  is also observed in the zoom-in Fermi surface (Fig.~3k, also see Supplementary Figure 8 and Supplementary Note 4), dispersion image (Fig.~3l) and corresponding MDCs (Fig.~3m) in the MicroARPES data. We have performed fine spatial scan on 5 different samples using MicroARPES where the domain size varies, however, the dispersions of individual domain remain the same (see Supplementary Figure 1, 2, 7 and Supplementary Note 1), suggesting that the measured dispersions are independent of the domain size. By combining NanoARPES and MicroARPES measurements, we reveal the electronic reconstructions of 1/6 and 1/3 \textbf{a}$^*$, which suggests that the two-fold symmetric electronic structure is likely associated with the 6$\times$1 reconstruction in the stripe phase. We note that in principle 5$\times$1 or 8$\times$1 would also be compatible with the two-fold symmetric Fermi surface, however signatures of 5$\times$1 or 8$\times$1 replicas have not been resolved experimentally, suggesting that those domains do not have significant contribution to the dispersions.

\begin{figure*}[htbp]
	\centering
	\includegraphics[width=16.5 cm]{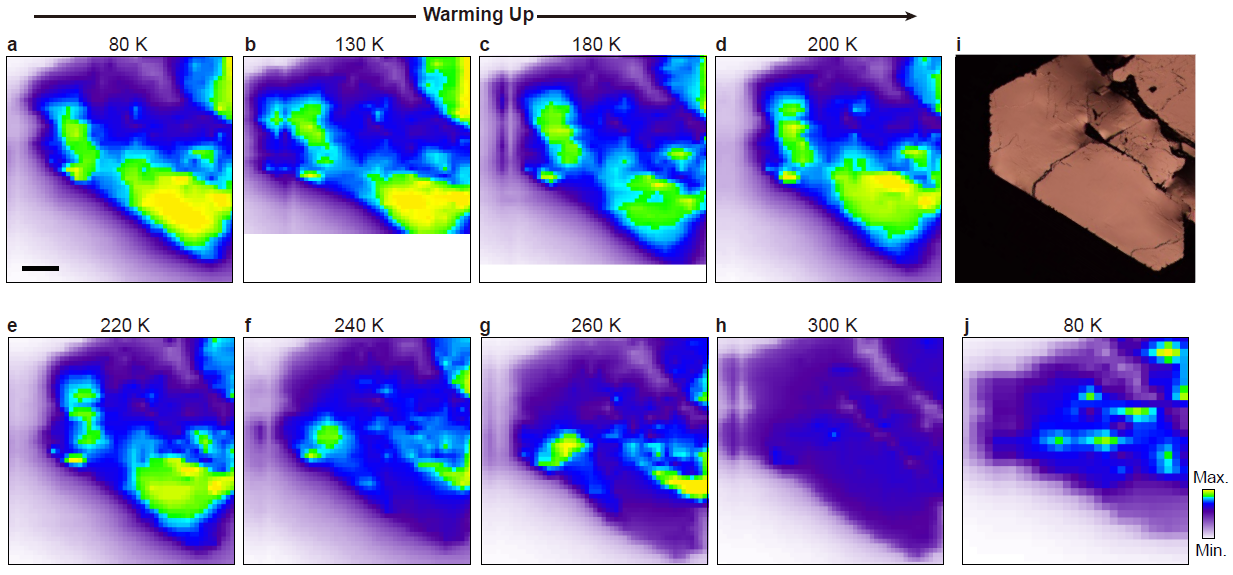}
	\caption{\textbf{Spatially-resolved MicroARPES intensity maps reveal the temperature dependence of the separated domains measured with laser source at photon energy of 6.2 eV.} \textbf{a-h} The evolution of spatially-resolved MicroARPES intensity maps upon warming by integrating over momentum range of $k_y$=(-0.1 $\rm\AA^{-1}$,0.1 $\rm \AA^{-1}$) and energy range of $E$=(-0.05 $\rm eV$,0). The scale bar is 200 $\mu$m. \textbf{i} Optical image of the sample. \textbf{j} Spatially-resolved MicroARPES intensity map at 80 K after cooling back.}
	\label{Fig3}
\end{figure*}

\textbf{Temperature evolution of the spatially-resolved intensity map.} To confirm that such spatial inhomogeneity is directly related to the stripe phase transition, we perform temperature-dependent MicroARPES measurement. Figure 4a-h shows spatially-resolved intensity maps measured at temperatures from 80 K to 300 K. Separated domains on the order of tens to hundreds of micrometers with different intensity contrast are clearly observed below the stripe transition temperature. Remarkably, above the stripe phase transition temperature, the spatial intensity map becomes much more homogeneous at 300 K (Fig.~4h) similar to its optical image. After cooling back to 80 K, the spatial inhomogeneity appears again but with different distribution, suggesting that its distribution is related to history (Fig.~4j). The observation of the stripe domains and its disappearance at high temperature provides direct evidence that the inhomogeneous electronic structure is an intrinsic property of the low temperature stripe phase.

\begin{figure*}[htbp]	
	\centering
	\includegraphics[width=16.5 cm]{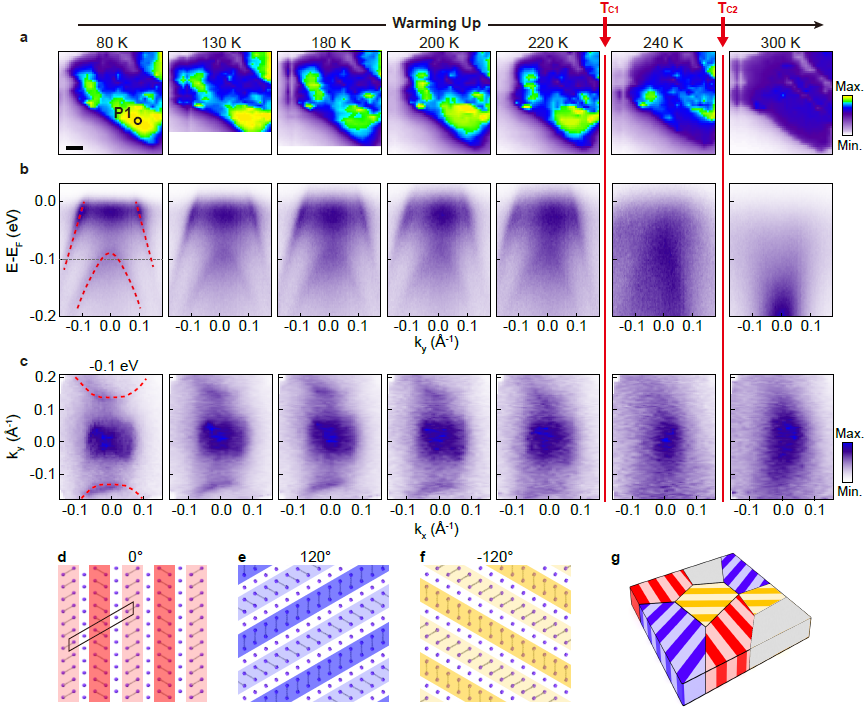}
	\caption{\textbf{Evolution of the electronic structure with temperature measured at a photon energy of 6.2 eV.} \textbf{a} Evolution of the spatial image during warming by integrating over momentum range of $k_y$=(-0.1 $\rm \AA^{-1}$,0.1 $\rm \AA^{-1}$) and energy range of $E$=(-0.05 $\rm eV$,0). The scale bar is 200 $\mu$m. \textbf{b} Evolution of the dispersion at k$_x$=0 during warming.  \textbf{c} Evolution of intensity maps at -0.1 eV during warming. \textbf{d-f} Three equivalent orientations of the stripes with 6$\times$1 reconstruction in the ab plane with dimers formed by Ir atoms. \textbf{g} A schematic drawing showing a mixture of 6$\times$1 stripe regions with different orientations (red, blue and yellow) and other regions (gray area).}
	\label{Fig4}
\end{figure*}

Figure 5a-c further shows the temperature evolution of the dispersion and intensity map measured in a single domain. Sharp dispersions near the $\bar{\Gamma}$ point are observed at low temperature and they disappear at 240 K and above. As was discussed above, the sharp dispersions are associated with the 6$\times$1 reconstruction, and their disappearance indicates a phase transition from 6$\times$1 to other reconstructions at T$_{\rm c1}$\cite{Plummer2017}. Further warming leads to another transition at T$_{\rm c2}$ near 300 K, which corresponds to the transition to 1$\times$1 phase. Similar evolution of the Fermi surface maps and dispersions is also observed for other domains but with a rotation angle of 120$^\circ$ (see Supplementary Figure 3 and Supplementary Note 2). In addition, broad dispersions and Fermi surface maps are also observed in some other locations (see Supplementary Figure 4), suggesting that there are also other regions in addition to the 6$\times$1 stripe, which is possibly caused by the small percentage of coexisting 8$\times$1 domain as revealed in LEED measurement shown in Supplementary Figure 5. Therefore, temperature dependent MicroARPES measurements show that the spatial inhomogeneity is directly related to the different orientations of the stripe phases, and there is a coexistence of both 6$\times$1 stripe domains with different stripe orientations (Fig.~5d-g) and other mixed domains (gray area in Fig.~5g) as schematically shown in Fig.~5g.

\section*{Conclusion}

In summary, the energy-, momentum- and space-resolving capability of NanoARPES and MicroARPES allows to visualize the separated domains and reveal the intrinsic and inhomogeneous electronic structure in the stripe phase  of IrTe$_2$.  Replica bands with 1/6 \textbf{a}$^*$ wave vector (or 6$\times$1 modulation) are identified in the dispersion which resembles the 6$\times$1 reconstruction of the high temperature electronic state (see Supplementary Figure 6 and Supplementary Note 3). We note that the period of the phase is strongly related to the Ir-Ir dimer concentration, and different dimer concentration leads to complex (3\textit{n}+2)$\times$1 stripe period. At the highest dimer concentration\cite{PanMHSTM2014, Kiryukhin2014}, this corresponds to the 6$\times$1 electronic ground state (Fig.~5d-f). Here by directly revealing the electronic structure of each individual domain using NanoARPES and MicroARPES, we show that the 6$\times$1 stripe phase is indeed the electronic ground state and the 6$\times$1 modulation is directly related to the stripe phase transition of IrTe$_2$. Our work resolves the puzzle in the electronic structure of the stripe phase of IrTe$_2$, and we envision that the application of NanoARPES and MicroARPES to other phase-separated systems can yield important information on the intrinsic underlying physics.

\begin{methods}

\subsection{Sample growth.}
High quality IrTe$_2$ single crystal was grown by self-flux method. Ir pellet (99.95\%, Alfa Aesar) and Te ingot (99.99\%, Alfa Aesar) in an atomic ratio of 5:95 were mixed together and sealed in an evacuated silica ampoule. The mixture was heated up to 900 $^{\circ}$C first and kept for several hours, then to 1150 $^{\circ}$C and kept for two days, finally cooled down to 920 $^{\circ}$C in several hours with a low rate. Liquid Te was separated from IrTe$_2$ single crystal by centrifugation.

\subsection{ARPES measurement.}
MicroARPES measurements have been performed in the home laboratory at Tsinghua University with fourth harmonic generation light source. The photon energy is set to 6.2 eV with $p$-polarization. The energy resolution was set to 15 meV. The beam size is 15 $\mu$m. The sample was measured in a working vacuum at greater than $7\times10^{-11}$ Torr.
Surface sensitive NanoARPES measurements were performed at the beamline ANTARES of the synchrotron SOLEIL\cite{AsensioNanoARPES} at France with a beam size of ~150 nm. The photon energy is 100 eV. The energy and angular resolution were set to 25 meV and 0.1 deg, respectively.

\end{methods}



\section*{References}

\begin{thebibliography}{10}
	\expandafter\ifx\csname url\endcsname\relax
	\def\url#1{\texttt{#1}}\fi
	\expandafter\ifx\csname urlprefix\endcsname\relax\def\urlprefix{URL }\fi
	\providecommand{\bibinfo}[2]{#2}
	\providecommand{\eprint}[2][]{\url{#2}}
	
	\bibitem{Elettra}
	\bibinfo{author}{Dudin, P.} \emph{et~al.}
	\newblock \bibinfo{title}{{Angle-resolved photoemission spectroscopy and
			imaging with a submicrometre probe at the {SPECTROMICROSCOPY-3.2L} beamline
			of {E}lettra}}.
	\newblock \emph{\bibinfo{journal}{J. Synchrotron Radiat.}}
	\textbf{\bibinfo{volume}{17}}, \bibinfo{pages}{445--450}
	(\bibinfo{year}{2010}).
	
	\bibitem{EliNanoARPES}
	\bibinfo{author}{Bostwick, A.}, \bibinfo{author}{Rotenberg, E.},
	\bibinfo{author}{Avila, J.} \& \bibinfo{author}{Asensio, M.~C.}
	\newblock \bibinfo{title}{Zooming in on electronic structure: {N}ano{ARPES} at
		{SOLEIL} and {ALS}}.
	\newblock \emph{\bibinfo{journal}{Synchrotron Radiat. News}}
	\textbf{\bibinfo{volume}{25}}, \bibinfo{pages}{19--25}
	(\bibinfo{year}{2012}).
	
	\bibitem{AsensioNanoARPES}
	\bibinfo{author}{Avila, J.} \& \bibinfo{author}{Asensio, M.~C.}
	\newblock \bibinfo{title}{First {NanoARPES} user facility available at
		{SOLEIL}: An innovative and powerful tool for studying advanced materials}.
	\newblock \emph{\bibinfo{journal}{Synchrotron Radiat. News}}
	\textbf{\bibinfo{volume}{27}}, \bibinfo{pages}{24--30}
	(\bibinfo{year}{2014}).
	
	\bibitem{ShimadaMARPES}
	\bibinfo{author}{Schwirr, E.} \emph{et~al.}
	\newblock \bibinfo{title}{Applications for ultimate spatial resolution in
		{LASER} based $\mu$-{ARPES}: a {F}e{S}e case study}.
	\newblock \emph{\bibinfo{journal}{AIP Conf. Proc.}}
	\textbf{\bibinfo{volume}{2054}}, \bibinfo{pages}{040017}
	(\bibinfo{year}{2019}).
	
	\bibitem{CobdenWSe2}
	\bibinfo{author}{Wilson, N.~R.} \emph{et~al.}
	\newblock \bibinfo{title}{Determination of band offsets, hybridization, and
		exciton binding in 2{D} semiconductor heterostructures}.
	\newblock \emph{\bibinfo{journal}{Sci. Adv.}} \textbf{\bibinfo{volume}{3}},
	\bibinfo{pages}{1601832} (\bibinfo{year}{2017}).
	
	\bibitem{MoTe2NanoARPES}
	\bibinfo{author}{Zhang, H.} \emph{et~al.}
	\newblock \bibinfo{title}{Resolving deep quantum-well states in atomically thin
		2{H}-{M}o{T}e$_2$ flakes by nanospot angle-resolved photoemission
		spectroscopy}.
	\newblock \emph{\bibinfo{journal}{Nano Lett.}} \textbf{\bibinfo{volume}{18}},
	\bibinfo{pages}{4664--4668} (\bibinfo{year}{2018}).
	
	\bibitem{JozwiakNP}
	\bibinfo{author}{Katoch, J.} \emph{et~al.}
	\newblock \bibinfo{title}{Giant spin-splitting and gap renormalization driven
		by trions in single-layer {WS}$_2$/h-{BN} heterostructures}.
	\newblock \emph{\bibinfo{journal}{Nat. Phys.}} \textbf{\bibinfo{volume}{14}},
	\bibinfo{pages}{355--359} (\bibinfo{year}{2018}).
	
	\bibitem{TrilayerG}
	\bibinfo{author}{Bao, C.} \emph{et~al.}
	\newblock \bibinfo{title}{Stacking-dependent electronic structure of trilayer
		graphene resolved by nanospot angle-resolved photoemission spectroscopy}.
	\newblock \emph{\bibinfo{journal}{Nano Lett.}} \textbf{\bibinfo{volume}{17}},
	\bibinfo{pages}{1564--1568} (\bibinfo{year}{2017}).
	
	\bibitem{tBLGPNAS}
	\bibinfo{author}{Yao, W.} \emph{et~al.}
	\newblock \bibinfo{title}{Quasicrystalline 30$^\circ$ twisted bilayer graphene
		as an incommensurate superlattice with strong interlayer coupling}.
	\newblock \emph{\bibinfo{journal}{Proc. Natl. Acad. Sci.}}
	\textbf{\bibinfo{volume}{115}}, \bibinfo{pages}{6928--6933}
	(\bibinfo{year}{2018}).
	
	\bibitem{CheongNat1999}
	\bibinfo{author}{Uehara, M.}, \bibinfo{author}{Mori, S.},
	\bibinfo{author}{Chen, C.} \& \bibinfo{author}{Cheong, S.}
	\newblock \bibinfo{title}{Percolate phase separation underlies colossal
		magnetoresistance in mixed-valence manganites}.
	\newblock \emph{\bibinfo{journal}{Nature}} \textbf{\bibinfo{volume}{399}},
	\bibinfo{pages}{560} (\bibinfo{year}{1999}).
	
	\bibitem{DavisNat2001}
	\bibinfo{author}{Pan, S.~H.} \emph{et~al.}
	\newblock \bibinfo{title}{Microscopic electronic inhomogeneity in the
		high-{T}$_c$ superconductor {B}i$_2${S}r$_2${C}a{C}u$_2${O}$_{8+x}$}.
	\newblock \emph{\bibinfo{journal}{Nature}} \textbf{\bibinfo{volume}{413}},
	\bibinfo{pages}{282} (\bibinfo{year}{2001}).
	
	\bibitem{DagottoSci2005}
	\bibinfo{author}{Dagotto, E.}
	\newblock \bibinfo{title}{Complexity in strongly correlated electronic
		systems}.
	\newblock \emph{\bibinfo{journal}{Science}} \textbf{\bibinfo{volume}{309}},
	\bibinfo{pages}{257} (\bibinfo{year}{2005}).
	
	\bibitem{BasovSci2007}
	\bibinfo{author}{Qazilbash, M.} \emph{et~al.}
	\newblock \bibinfo{title}{Mott transition in {VO}$_2$ revealed by infrared
		spectroscopy and nano-imaging}.
	\newblock \emph{\bibinfo{journal}{Science}} \textbf{\bibinfo{volume}{318}},
	\bibinfo{pages}{1750} (\bibinfo{year}{2007}).
	
	\bibitem{ShenSci2001}
	\bibinfo{author}{Lai, K.} \emph{et~al.}
	\newblock \bibinfo{title}{Mesoscopic percolating resistance network in a
		strained manganite thin film}.
	\newblock \emph{\bibinfo{journal}{Science}} \textbf{\bibinfo{volume}{329}},
	\bibinfo{pages}{190} (\bibinfo{year}{2010}).
	
	\bibitem{Kondo2020}
	\bibinfo{author}{Kuroda, K.} \emph{et~al.}
	\newblock \bibinfo{title}{Devil's staircase transition of the electronic
		structures in {CeSb}}.
	\newblock \emph{\bibinfo{journal}{Nat. Commun.}} \textbf{\bibinfo{volume}{11}},
	\bibinfo{pages}{2888} (\bibinfo{year}{2020}).
	
	\bibitem{Kim2019}
	\bibinfo{author}{Watson, M.~D.} \emph{et~al.}
	\newblock \bibinfo{title}{Probing the reconstructed {F}ermi surface of
		antiferromagnetic {BaFe}$_2${As}$_2$ in one domain}.
	\newblock \emph{\bibinfo{journal}{npj Quantum Materials}}
	\textbf{\bibinfo{volume}{4}}, \bibinfo{pages}{36} (\bibinfo{year}{2019}).
	
	\bibitem{Lanzara2020}
	\bibinfo{author}{Ma, J.} \emph{et~al.}
	\newblock \bibinfo{title}{Spatial nematic fluctuation in
		{BaFe}$_2$({As}$_{1-x}${P}$_x$)$_2$ revealed by spatially and angle-resolved
		photoemission spectroscopy}.
	\newblock \emph{\bibinfo{journal}{Phys. Rev. B}}
	\textbf{\bibinfo{volume}{101}}, \bibinfo{pages}{094515}
	(\bibinfo{year}{2020}).
	
	\bibitem{Kim2020}
	\bibinfo{author}{Rhodes, L.~C.}, \bibinfo{author}{Watson, M.~D.},
	\bibinfo{author}{Haghighirad, A.~A.}, \bibinfo{author}{Evtushinsky, D.~V.} \&
	\bibinfo{author}{Kim, T.~K.}
	\newblock \bibinfo{title}{Revealing the single electron pocket of {F}e{S}e in a
		single orthorhombic domain}.
	\newblock \emph{\bibinfo{journal}{Phys. Rev. B}}
	\textbf{\bibinfo{volume}{101}}, \bibinfo{pages}{235128}
	(\bibinfo{year}{2020}).
	
	\bibitem{KFeSe}
	\bibinfo{author}{Chen, Y.} \emph{et~al.}
	\newblock \bibinfo{title}{{Visualization of the electronic phase separation in
			superconducting K$_x$Fe$_{2-y}$Se$_2$}}.
	\newblock \emph{\bibinfo{journal}{Nano Research}}
	\textbf{\bibinfo{volume}{14}}, \bibinfo{pages}{823--828}
	(\bibinfo{year}{2021}).
	
	\bibitem{Cheong2012}
	\bibinfo{author}{Yang, J.~J.} \emph{et~al.}
	\newblock \bibinfo{title}{Charge-orbital density wave and superconductivity in
		the strong spin-orbit coupled {IrTe}$_{2}$:{Pd}}.
	\newblock \emph{\bibinfo{journal}{Phys. Rev. Lett.}}
	\textbf{\bibinfo{volume}{108}}, \bibinfo{pages}{116402}
	(\bibinfo{year}{2012}).
	
	\bibitem{Kiryukhin2014}
	\bibinfo{author}{Pascut, G.~L.} \emph{et~al.}
	\newblock \bibinfo{title}{Series of alternating states with unpolarized and
		spin-polarized bands in dimerized {IrTe}$_{2}$}.
	\newblock \emph{\bibinfo{journal}{Phys. Rev. B}} \textbf{\bibinfo{volume}{90}},
	\bibinfo{pages}{195122} (\bibinfo{year}{2014}).
	
	\bibitem{WuSTMPRL2013}
	\bibinfo{author}{Hsu, P.-J.} \emph{et~al.}
	\newblock \bibinfo{title}{Hysteretic melting transition of a soliton lattice in
		a commensurate charge modulation}.
	\newblock \emph{\bibinfo{journal}{Phys. Rev. Lett.}}
	\textbf{\bibinfo{volume}{111}}, \bibinfo{pages}{266401}
	(\bibinfo{year}{2013}).
	
	\bibitem{Plummer2017}
	\bibinfo{author}{Chen, C.} \emph{et~al.}
	\newblock \bibinfo{title}{Surface phases of the transition-metal dichalcogenide
		{IrTe}$_{2}$}.
	\newblock \emph{\bibinfo{journal}{Phys. Rev. B}} \textbf{\bibinfo{volume}{95}},
	\bibinfo{pages}{094118} (\bibinfo{year}{2017}).
	
	\bibitem{Monney2021}
	\bibinfo{author}{Nicholson, C.~W.} \emph{et~al.}
	\newblock \bibinfo{title}{Uniaxial strain-induced phase transition in the 2{D}
		topological semimetal {IrTe}$_2$}.
	\newblock \emph{\bibinfo{journal}{Commun. Mater.}}
	\textbf{\bibinfo{volume}{2}}, \bibinfo{pages}{1--8} (\bibinfo{year}{2021}).
	
	\bibitem{Ishiwata2013}
	\bibinfo{author}{Kamitani, M.} \emph{et~al.}
	\newblock \bibinfo{title}{Superconductivity in {Cu}${}_{x}${IrTe}${}_{2}$
		driven by interlayer hybridization}.
	\newblock \emph{\bibinfo{journal}{Phys. Rev. B}} \textbf{\bibinfo{volume}{87}},
	\bibinfo{pages}{180501} (\bibinfo{year}{2013}).
	
	\bibitem{DH2014}
	\bibinfo{author}{Qian, T.} \emph{et~al.}
	\newblock \bibinfo{title}{Structural phase transition associated with van
		{H}ove singularity in 5d transition metal compound {IrTe}$_2$}.
	\newblock \emph{\bibinfo{journal}{New J. Phys.}} \textbf{\bibinfo{volume}{16}},
	\bibinfo{pages}{123038} (\bibinfo{year}{2014}).
	
	\bibitem{Takashi2012}
	\bibinfo{author}{Ootsuki, D.} \emph{et~al.}
	\newblock \bibinfo{title}{Orbital degeneracy and peierls instability in the
		triangular-lattice superconductor
		{Ir}${}_{1\ensuremath{-}x}${Pt}${}_{x}${Te}${}_{2}$}.
	\newblock \emph{\bibinfo{journal}{Phys. Rev. B}} \textbf{\bibinfo{volume}{86}},
	\bibinfo{pages}{014519} (\bibinfo{year}{2012}).
	
	\bibitem{PascutPRL2014}
	\bibinfo{author}{Pascut, G.~L.} \emph{et~al.}
	\newblock \bibinfo{title}{Dimerization-induced cross-layer
		quasi-two-dimensionality in metallic {IrTe}$_{2}$}.
	\newblock \emph{\bibinfo{journal}{Phys. Rev. Lett.}}
	\textbf{\bibinfo{volume}{112}}, \bibinfo{pages}{086402}
	(\bibinfo{year}{2014}).
	
	\bibitem{KimCheong2014}
	\bibinfo{author}{Eom, M.~J.} \emph{et~al.}
	\newblock \bibinfo{title}{Dimerization-induced {F}ermi-surface reconstruction
		in {IrTe}$_2$}.
	\newblock \emph{\bibinfo{journal}{Phys. Rev. Lett.}}
	\textbf{\bibinfo{volume}{113}}, \bibinfo{pages}{266406}
	(\bibinfo{year}{2014}).
	
	\bibitem{David2013}
	\bibinfo{author}{Cao, H.} \emph{et~al.}
	\newblock \bibinfo{title}{Origin of the phase transition in {IrTe}${}_{2}$:
		Structural modulation and local bonding instability}.
	\newblock \emph{\bibinfo{journal}{Phys. Rev. B}} \textbf{\bibinfo{volume}{88}},
	\bibinfo{pages}{115122} (\bibinfo{year}{2013}).
	
	\bibitem{Artyukhin2020}
	\bibinfo{author}{Saleh, G.} \& \bibinfo{author}{Artyukhin, S.}
	\newblock \bibinfo{title}{First-principles theory of phase transitions in
		{IrTe}$_2$}.
	\newblock \emph{\bibinfo{journal}{J. Phys. Chem. Lett.}}
	\textbf{\bibinfo{volume}{11}}, \bibinfo{pages}{2127--2132}
	(\bibinfo{year}{2020}).
	
	\bibitem{Min2015}
	\bibinfo{author}{Kim, K.} \emph{et~al.}
	\newblock \bibinfo{title}{Origin of first-order-type electronic and structural
		transitions in {IrTe}$_{2}$}.
	\newblock \emph{\bibinfo{journal}{Phys. Rev. Lett.}}
	\textbf{\bibinfo{volume}{114}}, \bibinfo{pages}{136401}
	(\bibinfo{year}{2015}).
	
	\bibitem{ARPES2013}
	\bibinfo{author}{Ootsuki, D.} \emph{et~al.}
	\newblock \bibinfo{title}{Electronic structure reconstruction by orbital
		symmetry breaking in {IrTe}$_2$}.
	\newblock \emph{\bibinfo{journal}{J. Phys. Soc. Jpn.}}
	\textbf{\bibinfo{volume}{82}}, \bibinfo{pages}{093704}
	(\bibinfo{year}{2013}).
	
	\bibitem{Park2017}
	\bibinfo{author}{Lee, H.} \emph{et~al.}
	\newblock \bibinfo{title}{Electronic reconstruction on dimerized {IrTe}$_2$}.
	\newblock \emph{\bibinfo{journal}{Europhys. Lett.}}
	\textbf{\bibinfo{volume}{120}}, \bibinfo{pages}{47003}
	(\bibinfo{year}{2017}).
	
	\bibitem{ARPES2017}
	\bibinfo{author}{Ootsuki, D.} \emph{et~al.}
	\newblock \bibinfo{title}{A novel one-dimensional electronic state at
		{IrTe}$_2$ surface}.
	\newblock \emph{\bibinfo{journal}{J. Phys. Soc. Jpn.}}
	\textbf{\bibinfo{volume}{86}}, \bibinfo{pages}{123704}
	(\bibinfo{year}{2017}).
	
	\bibitem{OOTSUKI2018}
	\bibinfo{title}{Interplay between spin-orbit interaction and stripe-type
		charge-orbital order of {IrTe}$_2$}.
	\newblock \emph{\bibinfo{journal}{J. Phys. Chem. Solids}}
	\bibinfo{pages}{270--274} (\bibinfo{year}{2018}).
	
	\bibitem{HengsbergerPRB2018}
	\bibinfo{author}{Monney, C.} \emph{et~al.}
	\newblock \bibinfo{title}{Robustness of the charge-ordered phases in
		{I}r{T}e$_2$ against photoexcitation}.
	\newblock \emph{\bibinfo{journal}{Phys. Rev. B}} \textbf{\bibinfo{volume}{97}},
	\bibinfo{pages}{2} (\bibinfo{year}{2018}).
	
	\bibitem{MonneyPRB2020}
	\bibinfo{author}{Rumo, M.} \emph{et~al.}
	\newblock \bibinfo{title}{Examining the surface phase diagram of {I}r{T}e$_2$
		with photoemission}.
	\newblock \emph{\bibinfo{journal}{Phys. Rev. B}}
	\textbf{\bibinfo{volume}{101}}, \bibinfo{pages}{235120}
	(\bibinfo{year}{2020}).
	
	\bibitem{PanMHSTM2014}
	\bibinfo{author}{Li, Q.} \emph{et~al.}
	\newblock \bibinfo{title}{Bond competition and phase evolution on the
		{IrTe}$_2$ surface}.
	\newblock \emph{\bibinfo{journal}{Nat. Commun.}} \textbf{\bibinfo{volume}{5}},
	\bibinfo{pages}{5358} (\bibinfo{year}{2014}).
	
\end{thebibliography}


\begin{addendum}
 \item This work is supported by National Natural Science Foundation of China (Grants No.~11725418 and 11427903), the National Key R\&D Program of China (Grants No.~2020YFA0308800, 2016YFA0301004), Tsinghua University Initiative Scientific Research Program and Tohoku-Tsinghua Collaborative Research Fund, and Beijing Advanced Innovation Center for Future Chip (ICFC). We acknowledge SOLEIL for provision of synchrotron radiation facilities.
 
 \item[Author Contributions] Shuyun Z. conceived the research project. C.B. and Hongyun Z. performed the NanoARPES measurements and analyzed the data. C.B.,  Q.L., Shaohua Z., L.L., K.D. and W.Y. performed the Laser-based Micro-ARPES measurements and analyzed the data. Haoxiong Z., K.Z. and Y.W. grew and characterized the samples. C.C., J.A. and M.C.A. provided support for the NanoARPES experiments. C.B. and Shuyun Z. wrote the manuscript, and all authors commented on the manuscript.

 \item[Competing Interests] All other authors declare no competing interests.

  \item[Data availability] The data that supports the findings of this study are available within the article.
\end{addendum}


\end{document}